\documentstyle[12pt,equations]{article}
\newcommand{\ar}{\alpha}
\newcommand{\bb}{\beta}
\newcommand{\gm}{\gamma}

\newcommand{\en}{\epsilon}
\newcommand{\dd}{\delta}
\newcommand{\ld}{\lambda}

\newcommand{\be}{\begin{equation}}
\newcommand{\ee}{\end{equation}}
\newcommand{\bea}{\begin{eqnarray}}
\newcommand{\eea}{\end{eqnarray}}
\newcommand{\nn}{\nonumber}
\newcommand{\vf}{\varphi}
\newcommand{\bse}{\begin{subequations}}
\newcommand{\ese}{\end{subequations}}
\def\theequation{\arabic{section}.\arabic{equation}}
\begin{document}
%
%
%\begin{flushright}
%{\large INS \# 193/92}
%\end{flushright}
%

\begin{center}
{\Large {\bf INTEGRABLE QUANTUM MAPPINGS AND \\
QUANTIZATION ASPECTS OF INTEGRABLE DISCRETE-TIME SYSTEMS}}$^{\ast}$\\
\end{center}
\vspace{.4in}

\footnotetext{$^{\ast}$ Two talks presented by both authors at the
NATO Advanced Research Workshop on {\em Applications of Analytic
and Geometric Methods to Nonlinear Differential Equations}, Exeter, July 1992.}

\begin{center}
F.W. Nijhoff$^{\dagger}$\footnote{$^{\dagger}$
{\em Department of Mathematics and Computer Science\\
and Institute for Nonlinear Studies,\\
Clarkson University, Potsdam NY 13699-5815, USA}}
{\small and} H.W. Capel$^{\ddag}$\footnote{ $^{\ddag}$
{\em Institute for Theoretical Physics, University of Amsterdam,\\
Valckenierstraat 65, 1018 XE Amsterdam, The Netherlands}}
\end{center}
\vspace{.5in}

\centerline{ {\bf Abstract} }
\vspace{.2in}

{\small We study a quantum Yang-Baxter structure associated with
non-ultralocal lattice models. We discuss the canonical structure of a class
of integrable quantum mappings, i.e. canonical transformations preserving the
basic commutation relations. As a particular class of solutions we present two
examples of quantum mappings associated with the lattice analogues of the KdV
and MKdV equations, together with their exact quantum invariants. }

\vspace{.5in}

\addtocounter{footnote}{-2}

\section{Introduction}
\setcounter{equation}{0}

\noindent
Discrete integrable models, in which the spatial dimension is discretized, but
the time is continuous, have traditionally played an important role in
mathematics and physics, both in the classical as well as in the quantum
regime. On the quantum level the algebraic structure of integrable systems is
discussed in terms of quantum groups \cite{1}-\cite{3}. The discretized
version of such models has played a particular role in this respect, e.g. in
the quantum inverse scattering method \cite{4}. The models, in which also the
time-flow is discretized (i.e. integrable lattices or partial difference
equations), have been considered on the classical level in a number of papers
 \cite{5,6}. Recently, they have become of interest in connection with the
construction of integrable {\em mappings}, i.e. finite-dimensional reductions
of these integrable lattice equations, \cite{7,8}. Their integrability is to
be understood in the sense that the discrete time-flow is the iterate of a
canonical transformation, preserving a suitable symplectic structure, leading
to invariants which are in involution with respect to this symplectic form. A
theorem \`a la Liouville then tells us in analogy with the continuous-time
situation that one can linearize the discrete-time flow on a hypertorus which
is the intersection of the level sets of the invariants, \cite{9}. Integrable
mappings have been considered from a slightly different perspective also in
the recent literature, cf. e.g. \cite{10}-\cite{13}.

Integrable two-dimensional lattices arise, both on the classical as well as on
the quantum level, as the compatibility conditions of a discrete-time ZS
(Zakharov-Shabat) system
\be
L'_n(\lambda ) \cdot M_n(\lambda )\ =\ M_{n+1}(\lambda ) \cdot L_n(\lambda ) \
 ,
\ee
in which $\lambda$ is a spectral parameter, $L_n$ is the lattice translation
operator at site $n$, and the prime denotes the discrete time-shift
corresponding to a translation in the second lattice direction. As $L$ and
$M$, in the quantum case, depend on operators, the question of operator
ordering becomes important. Throughout this paper we impose in the quantum
case as a normal order the order which is induced by the lattice enumeration,
with $n$ increasing from the left to the right. Finite-dimensional mappings
are obtained from (1.1) imposing a periodicity condition
\be
L_n (\lambda )\  =\  L_{n+P} (\lambda ) \ \  , \ \  M_n (\lambda ) \  =\
M_{n+P} (\lambda )
\ee
for some $P \in {\bf N}$.

In a recent paper \cite{14}, we introduced a novel quantum structure that is
appropriate for obtaining an integrable quantization of mappings of the
so-called KdV-type, i.e. mappings derived from a lattice version of the KdV
equation, cf. \cite{7}. In this paper we will review the construction of such
integrable quantum mappings and their quantum invariants and we consider also
the quantum mappings associated with the lattice version of the MKdV
equation. As was indicated in \cite{8}, it turns out that these mappings and
their underlying integrable lattices are --on the classical level-- symplectic
with respect to  a so-called {\em non-ultralocal\/} Poisson structure
 \cite{8},\cite{15}. In the continuous-time case such (classical)
non-ultralocal r-matrix structures have been studied in a number of papers,
cf. e.g. \cite{16}-\cite{20}. The discrete version of the non-ultralocal
Poisson bracket structure reads, cf. ref. \cite{20},
\bea
\{ L_{n,1} , L_{m,2} \} &=&
- \dd_{n,m+1}\,L_{n,1}\,s^+_{12}\,L_{m,2}\ +\
\dd_{n+1,m}\,L_{m,2}\,s^-_{12}\,L_{n,1}\nn \\
&+& \dd_{n,m} \left[ r^+_{12}\,L_{n,1}L_{m,2}\,
-\,L_{n,1} L_{m,2}\,r^-_{12} \right] \   ,
\eea
Throughout this paper we adopt the usual convention that the subscripts $1,2,
\cdots$ in (1.3) denote the factors in a matricial tensor product,
i.e. $A_{i_1,i_2,\dots ,i_M}=A_{i_1,i_2,\dots ,i_M}(\lambda _1,\lambda _2,
\dots ,\lambda _M)$ denotes a matrix acting nontrivially only on the factors
labeled by $i_1,i_2,\dots ,i_M$ of a tensor product $\otimes_{\ar}\,V_{\ar}$,
of vector spaces $V_{\ar}$ and trivially on the other factors, cf.\@ also
e.g.\@ \cite{4,15}. For example, in eq. (1.3),  the subscripts $\ar,\bb=1,2,
\cdots$ for the operator matrices $L_{n,\ar}$ denote the corresponding factor
on which this $L_n$ acts (acting trivially on the other factors), i.e. $L_{n,
1} = L_n(\lambda_1 ) \otimes {\bf 1}$, $L_{n,2}={\bf 1}\otimes L_n(\lambda_2
)$. We suppress the explicit dependence on the spectral parameter $\lambda
=\lambda_1$ respectively $\lambda = \lambda_2$, assuming that each value
accompanies its respective factor in the tensor product.

Eq. (1.3) defines a proper Poisson bracket provided that the
following relations hold for $s^{\pm}=s^{\pm}(\lambda _1,\lambda_2)$ and
$r^{\pm}=r^{\pm}(\lambda_1,\lambda_2)$:
\be
s^-_{12}(\lambda_1,\lambda_2)\ =\ s^+_{21}(\lambda_2,\lambda_1)  \    \ ,\   \
r^{\pm}_{12}(\lambda_1,\lambda_2)\ =\ - r^{\pm}_{21}(\lambda_2,\lambda_1) \  ,
\ee
to ensure the skew-symmetry, and
\bse
\bea
\left[ r_{12}^{\pm}\,,\,r_{13}^{\pm}\right]\ +\
\left[ r_{12}^{\pm}\,,\,r_{23}^{\pm}\right]\ +\
\left[ r_{13}^{\pm}\,,\,r_{23}^{\pm}\right] &=& 0\   , \label{eq:yb1} \\
\left[ s_{12}^{\pm}\,,\,s_{13}^{\pm}\right]\ +\
\left[ s_{12}^{\pm}\,,\,r_{23}^{\pm}\right]\ +\
\left[ s_{13}^{\pm}\,,\,r_{23}^{\pm}\right] &=& 0  , \label{eq:yb2}
\eea
\ese to ensure that the the Jacobi identities hold for the Poisson bracket
(1.3).  The relation (\ref{eq:yb1}  for $r^{\pm}$ is nothing but the usual
classical Yang-Baxter equation (CYBE). As a consequence of the ZS system (1.1)
we have on the classical level a complete family of invariants of the mapping
, namely by introducing the associated monodromy matrix $T(\lambda )$,
obtained by gluing the elementary translation matrices $L_j$ along a line
connecting the sites 1 and $P+1$ over one period $P$, namely
\be
T(\lambda )\ \equiv \stackrel{\longleftarrow}{\prod_{n=1}^P}\
L_n(\lambda )\ .
\ee
In order to be able to integrate (1.3) to obtain Poisson brackets for
the monodromy matrix we need in addition to these relations the extra relation
\be
r^+_{12}\,-\,s^+_{12}\ =\ r^-_{12}\,-\,s^-_{12}\   .
\ee
In the classical case the traces of powers of the monodromy matrix are
invariant under the mapping as a consequence of
\be
T' (\lambda ) = M_{P+1} (\lambda ) T(\lambda )  M^{-1}_1 (\lambda )
\ee
and the periodicity condition $M_{P+1}=M_1$, thus leading to a sufficent
number of invariants which are obtained by expanding the traces in powers of
the spectral parameter $\lambda$. The involution property of the classical
invariants follows from the Poisson bracket
\be
\{ {\rm tr} T(\lambda ) , {\rm tr} T(\lambda ') \} = 0
\ee
which can be derived from (1.3).

For the quantum mappings we will use the structure of \cite{14,21} which is
the quantum analogue of this non-ultralocal Poisson structure. In the
continuous-time case such a novel quantization scheme was proposed in ref.
\cite{22}, in connection with the quantum Toda theory. Similar structures with
continuous time flow have been introduced also for the quantum
Wess-Zumino-Novikov-Witten (WZNW) theory with discrete spatial variable, cf.
\cite{23,24}. When considering discrete-time flows some interesting new
features arise, as was indicated in \cite{14,21}. In fact, the conventional
point of view, that the $M$-part of the Lax equations does not need to be
considered explicitly in order to construct quantum invariants, is no longer
true. Therefore, one needs to establish the complete quantum algebra,
containing commutation relations between the $L$-operators as well as between
the $L$- and the $M$-operators, and between the $M$-operators themselves. As a
consequence we will find, in the quantum mappings under consideration,
non-trivial quantum corrections in the quantum invariants of the mappings.
 From an algebraic point of view, the basic algebraic relations for the
monodromy matrices, that are relevant in the context of non-ultralocal models,
are the algebras of currents introduced in various papers,
\cite{25}-\cite{27}, in different contexts. Interestingly enough, the
relations between the monodromy matrix and the time-part of the Lax
representation are very similar to the relations associated with the
description of the cotangent bundle of a quantum group $(T^{\ast}G)_q$
 \cite{28}.

The outline of this paper is as follows. In section 2 we introduce the basic
ingredients of the non-ultralocal quantum $R,S$-structure. In section 3 we
investigate the canonical structure of quantum gauge- or similarity
transformations,  leading to (integrable) quantum mappings. This leads to a
`full' Yang-Baxter structure including the discrete-time part of the Lax
representation. In section 4 we present two examples of this structure:
quantum mappings associated with the lattice KdV and with the lattice MKdV
equation. In order to establish the quantum integrability of these mappings,
we then develop in section 5 the `full' quantum structure for the  monodromy
matrix, and show how to construct commuting families of exact quantum
invariants for these mappings.

\section{Non-ultralocal Yang-Baxter structure}
\setcounter{equation}{0}

We now define a quantum Yang-Baxter structure that is adequate for the
mappings in this paper, i.e. discrete-time systems arising (both on the
classical as well as quantum level) from compatibility equations of the form
of (1.1).

We introduce the {\em quantum} $L$-operator $L_n(\lambda )$ at each site $n$ of
a one-dimensional lattice, which is a matrix whose entries are quantum
operators
(acting on some properly chosen Hilbert space). The operators $L_n(\lambda )$
are
supposed to have only non-trivial commutation relations between themselves on
the same and nearest-neighbour sites, namely as follows
\bse
\bea
R_{12}^+\,L_{n,1} \cdot\,L_{n,2} &=&
L_{n,2} \cdot\,L_{n,1}\,R_{12}^-\   \label{eq:RLL}  \\
L_{n+1,1} \cdot S^+_{12}\,L_{n,2} &=&
L_{n,2} \cdot\,L_{n+1,1}\   ,  \label{eq:LSL}  \\
L_{n,1} \cdot \, L_{m,2} &=& L_{m,2} \cdot \, L_{n,1} \, , \, \mid n-m \mid
\geq 2\    . \label{eq:2.1c}
\eea
\ese
These relations are the quantum analogue of the {\em non-ultralocal} Poisson
bracket (1.3). We will show in section 4 that the quantum mappings
provide examples of such a non-ultralocal quantum $R$-matrix structure.

The compatibility relations of the equations (\ref{eq:RLL})-(\ref{eq:2.1c})
lead
to the following consistency conditions on $R^{\pm}$ and $S$
\bse
\bea
R^{\pm}_{12}\,R^{\pm}_{13}\,R^{\pm}_{23} &=&
R^{\pm}_{23}\,R^{\pm}_{13}\,R^{\pm}_{12}\  ,  \label{eq:RRR} \\
R^{\pm}_{23}\,S^{\pm}_{12}\,S^{\pm}_{13} &=&
S^{\pm}_{13}\,S^{\pm}_{12}\,R^{\pm}_{23}\  , \label{eq:RSS}
\eea
\ese where $S^+_{12}=S^-_{21}$. Eq. (2.2a) is the quantum Yang-Baxter equation
(QYBE's) for $R^{\pm}$ coupled with an additional equation (2.2b) for
$S^{\pm}$. They were given for the first time explicitly in \cite{14}, but
they are implicit in the previous literature \cite{25}-\cite{27}, where the
relations (2.8) to be given below below were given in more special situations.
For a derivation of eqs. (2.2) see appendix A.

In order to establish that the structure given by the commutation relations
(2.1) allows for suitable commutation relations for the monodromy
matrix, we need to impose in addition to (2.2) that
\be
R_{12}^{\pm}\,S^{\pm}_{12}\ =\ S^{\mp}_{12}\,R_{12}^{\mp}\    .
\ee
Using these relations it is easy to establish that each sign of eqs.
(\ref{eq:RRR}),(\ref{eq:RSS}) can be combined into a single equation as
follows
\be
R^{\pm}_{12} \left( R^{\pm}_{13} S^{\pm}_{13} \right) S^{\pm}_{12}
\left( R^{\pm}_{23} S^{\pm}_{23} \right)\ =\
\left( R^{\pm}_{23} S^{\pm}_{23} \right) S^{\mp}_{12}
\left( R^{\pm}_{13} S^{\pm}_{13} \right) R^{\mp}_{12}\   .  \label{eq:RSSR}
\ee

At this point it is useful to introduce the following decomposition
of the monodromy matrix (1.6)
\be
T\ =\ T_n^+ \cdot T_n^-\  ,
\ee
in which
\be
T^+_n(\lambda )\ =\ \stackrel{\longleftarrow}{\prod_{j=n+1}^P}\, L_j(\lambda
)\    \ ,\    \ T^-_n(\lambda )\ =\ \stackrel{\longleftarrow}{\prod_{j=1}^n}\,
L_j(\lambda )\     ,
\ee
First, one derives for the monodromy matrices $T_n^+$ and $T_n^-$
the following set of relations
\bse
\bea
R^+_{12}\,T_{n,1}^{\pm} \cdot T_{n,2}^{\pm}\ =\ T_{n,2}^{\pm}
\cdot T_{n,1}^{\pm}\,R_{12}^-\  ,  \label{eq:2.7a} \\
T_{n,1}^+ \cdot S^+_{12}\,T_{n,2}^-\ =\
T_{n,2}^- \cdot S^-_{12}\,T_{n,1}^+\  ,   \label{eq:TST}
\eea
\ese
for $2\leq n\leq P-1$.

Next, taking into account the periodic boundary conditions we obtain for the
monodromy matrix the commutation relations
\be
R_{12}^{+}\,T_1 \cdot \,S^{+}_{12}\, T_2 \ =\ T_2 \cdot \, S^{-}_{12}\,
T_1\, R_{12}^{-}\    .
\ee
Some details of the derivation of eqs. (2.7), (2.8) are
presented in appendix B.\\

\noindent {\bf Remarks:}
\paragraph{a)} From the theory of quantum groups, \cite{3}, we know that the
relations (2.7) are the defining relations for a quasi-triangular
Hopf algebra ${\cal A}^{\pm}$ , generated by the entries of $T^{\pm}_n$ and
the unit, with defining relations (2.8). The co-algebra structure on
${\cal A}$ is defined with the following coproduct
\be
\Delta(T_n^{+})\ =\ T_n^{+} \stackrel{\longrightarrow}{\dot{\otimes}}
T^{+}_n\    \ ,\    \
\Delta(T_n^{-})\ =\ T_n^{-} \stackrel{\longleftarrow}{\dot{\otimes}}
T^{-}_n\    \ ,\    \
  \label{eq:Delta}
\ee
in which the dot denotes a matrix product. The algebra generated by the unit
and $T$ is no longer a Hopf algebra, but instead of that $T$ and the unit 1
generate a Hopf-ideal ${\cal A}$ \cite{27}
\be
\Delta({\cal A}) \subset {\cal A}^{\pm} \otimes {\cal A}  \  . \label{eq:2.10}
\ee
In fact, the coproduct on the generators $T$ is defined as
\be
\Delta(T)\ =\ \left( T^+\otimes {\bf 1}\right) \cdot \left( {\bf 1}\otimes
 T\right) \cdot \left( T^-\otimes {\bf 1}\right)\  , \label{eq:2.11}
\ee
as a consequence of (2.5) and (\ref{eq:Delta}).

\paragraph{b)} The classical limit of the quantum structure
(2.1)-(2.2) is easily obtained by considering the quasi-classical
expansion
\bea
S^{\pm}_{12} & = & {\bf 1}\otimes {\bf 1}\ -\ h\,s^{\pm}_{12}\ +\ {\cal
O}(h^2)\  , \nn \\
R^{\pm}_{12} & = & {\bf 1}\otimes {\bf 1}\ +\ h\,r^{\pm}_{12}\ +\ {\cal
O}(h^2)\
,  \label{eq:2.12}
\eea
In this limit the quantum commutation relations (2.1) yield the non-ultralocal
Poisson bracket structure given in eqs. (1.3)-(1.5).

\section{Quantum Mappings}
\setcounter{equation}{0}

We are interested in the canonical structure of discrete-time integrable
systems, i.e. systems for which the time evolution is given by an iteration of
mappings. If the mapping contains quantum operators, the commutation relations
with the monodromy or Lax matrices become nontrivial and it is not a priori
clear in this case that the Yang-Baxter structure is preserved. Furthermore,
the traces of powers of the monodromy matrix are no longer trivially invariant
as the cyclic property of the traces is no longer true for operator-valued
arguments.

Let us comment why in the discrete-time case the spatial part of the
Yang-Baxter equations is not sufficient for quantum integrable systems. When
dealing with Yang-Baxter structures for continuous-time systems, cf.\@ e.g.\@
\cite{4,15}, it is not necessary to consider explicitly the commutation
relations (Poisson brackets on the classical level) involving the
$M$-operator, i.e. the time-dependent part of the Lax pair. For instance,
knowing the commutation relations for the spatial part, i.e.\@ the
$L$-operator, allows one to construct all relevant objects, namely the
(conserved) Hamiltonians of the system as well as the corresponding
$M$-operators, cf. \cite{15} and \cite{29} for the quantum case. In doing so,
one may use the fact that the time evolution is governed by a canonical
transformation which leaves the commutation relations between operators
invariant. In other words one can reconstruct the allowed integrable
continuous-time flows. In the discrete-time case this is no longer true.
Firstly, on the lattice there is not a preferred `spatial' versus `time-like'
direction, as both components appear on the same footing, and hence, there is
no natural impetus to introduce reconstruction formulae expressing one of the
Lax operators in terms of the other one. Secondly, the process of integrating
an integrable continuous-time flow to a finite-step discrete-time flow is a
highly non-trivial procedure, in which -- a priori -- it is not at all clear
that the integrated time flow will lead to nice `local' partial difference
equations described by explicit closed-form expressions. Furthermore, in the
time-discrete case the symplectic property is not automatically guaranteed, it
is not  a priori clear that the commutation properties of operators are
invariant under the mapping. On the other hand, the symplectic property is a
necessary ingredient for mappings to be integrable \cite{9}.

However, the explicit investigations of mappings obtained by reduction from
integrable partial difference equations reveal that this is actually the case.
Probably there is a highly nonlinear self-consistent mechanism at work in
these systems. In order to get a grasp on this mechanism we feel that it is
necessary to take the $M$-part of the Lax or ZS system into account, and
investigate the {\em full} quantum structure involved in these systems,
consisting of commutation relations between the $L$-part as well as of the
$M$-part of the Lax pair.

In \cite{14} and \cite{21} we have introduced such a full Yang-Baxter
structure taking account of the spatial as well as the time part of the Lax
pair. The structure is obtained by supplying in addition to eqs.
(2.1) the following equations.
\bse
\bea
M_{n+1,1} \cdot S^+_{12} L_{n,2} &=& L_{n,2} \cdot M_{n+1,1}\   ,
\label{eq:LM1}  \\
L^{\prime}_{n,2} \cdot S^-_{12} M_{n,1} &=&
M_{n,1} \cdot L^{\prime}_{n,2}   , \label{eq:LM2}
\eea
\ese
and
\bse
\bea
R^+_{12}\,M_{n,1} \cdot M_{n,2} &=& M_{n,2} \cdot M_{n,1} R^-_{12} \  ,
\label{eq:3.2a} \\
M^{\prime}_{n,1} \cdot S^+_{12} M_{n,2} &=& M_{n,2} \cdot M^{\prime}_{n,1}\
{}.
\label{eq:RMM}
\eea
\ese
The trivial commutation relations are the following
\bse
\bea
M_{n,1} \cdot L_{m,2} &=& L_{m,2} \cdot M_{n,1}\    \ ,\    \  |n-m|\geq 2\
, \label{eq:3.3a}  \\
M_{n+1,1} \cdot L^{\prime}_{m,2} &=& L^{\prime}_{m,2} \cdot M_{n+1,1}\   \ ,\
 \ |n-m|\geq 2 \ , \label{eq:triv}   \\
M_{n,1} \cdot M_{m,2} &=& M_{m,2} \cdot  M_{n,1}\   \ ,\
 \ |n-m|\geq 2 \ ,  \label{eq:3.3c}
\eea
\ese
in combination with
\bse
\bea
M^{\prime\prime\dots }_{n,1} \cdot M_{n,2} &=&
M_{n,2} \cdot M^{\prime\prime\dots }_{n,1}\  ,  \label{eq:3.4a} \\
M^{\prime}_{n+1,1} \cdot M_{n,2} &=&
M_{n,2} \cdot M^{\prime}_{n+1,1}\  ,  \label{eq:3.4b}\\
M^{\prime}_{n+1,1} \cdot L_{n,2} &=&
L_{n,2} \cdot M^{\prime}_{n+1,1}\  ,    \label{eq:3.4c}
\eea
\ese
for multiple applications of the mapping. We shall not specify other
commutation relations, as they do not belong to the Yang-Baxter structure.
More precisely, one may notice in the explicit examples of section 4 that the
commutation relations
$$
[L_n\stackrel{\otimes}{,}M_n]\    \ ,\    \   [L_{n+1}\stackrel{\otimes}{,
}M_n]\    \ ,\    \  [M_{n+1}\stackrel{\otimes}{,}L^{\prime}_{n}]\    \ ,\
\  [M_{n+1}\stackrel{\otimes}{,}L^{\prime}_{n-1}]\    \ ,\    \
[M_{n+1}\stackrel{\otimes}{,}M_{n}]\   ,
$$
are nontrivial, and they depend on the details of the system satisfying the
Yang-Baxter equations. However, in order for the Yang-Baxter structure to be
preserved under the mapping, we do not need information on these latter
commutation relations. Let us now make a more explicit statement on the
invariance of the non-ultralocal Yang-Baxter system. The full Yang-Baxter
structure consists of two sets of nontrivial commutation relations
\begin{tabbing}
ii)ss \= \kill
i) \> Eqs. (\ref{eq:RLL}), (\ref{eq:LSL}) and (\ref{eq:LM1})\\
ii) \> Eqs. (\ref{eq:LM2}), (\ref{eq:3.2a}) and (\ref{eq:RMM}).
\end{tabbing}
Imposing the first set of equations for all $n$, and the second set of
equations for a fixed value of $n$ but for all iterates of the mapping and
using also trivial commutation relations, it follows that the first set of
equations, and in particular the commutation relations between the matrices
$L_n$, is invariant under the mapping
\be
L_n \rightarrow L_n^{\prime} = M_{n+1} \, L_n \, M_n^{-1} \  , \label{eq:3.5}
\ee
see appendix C for some details.

For the quantum mappings under consideration here the operator $L_n$ has a
composite structure, i.e.
\be
L_n = V_{2n} \cdot V_{2n-1} \label{eq:3.6}
\ee
and the commutation relations of the Yang-Baxter structure involving the $L_n$
can be inferred from the commutation relations among the $V_n$ themselves, as
well as the commutation relations between the $V_n$ and $M_m$. In fact,
imposing the commutation relations
\bse
\bea
V_{n+1,1}\cdot S^+_{12}(n)\,V_{n,2} &=& V_{n,2}\cdot V_{n+1,1}\  ,
\label{eq:VV} \\
R^+_{12}\,V_{n,1}\cdot V_{n,2} &=& V_{n,2}\cdot V_{n,1}\,R^-_{12}\  ,
\label{eq:3.7b} \\
V_{n,1}\cdot V_{m,2} &=& V_{m,2} \cdot V_{n,1}\, , \, |\, n-m |  \geq 2 \
,\label{eq:3.7c}
\eea
\ese
we obtain the relations (2.1) as can be easily verified.

In eq. (\ref{eq:VV}) the $S^+_{12} (2n)$ is independent of $n$ and is equal to
the $S^+_{12}$ occuring in eq. (2.1b). For odd values of $n$, $S^+_{12}
(n)$ may be a different solution of eqs. (2.1)-(2.3). The proof of
eq. (\ref{eq:RLL}) from eq. (\ref{eq:3.7b}) is essentially the same as the
proof
in appendix B showing how eq. (\ref{eq:2.7a}) is obtained from eqs.
(2.1).\\

Next we impose the commutation relations between the operators $V_n$ and
$M_n$. The only nonvanishing commutation relations involving $M_n$ are taken
to be the following ones
\be
\begin{array}{l} V_{2n+1}\\V_{2n}\\V_{2n-1}\\V_{2n-2} \end{array}
\leftrightarrow M_n \leftrightarrow
\begin{array}{l} V^{\prime}_{2n-1}\\V^{\prime}_{2n-2}\\V^{\prime}_{2n-3}\\
V^{\prime}_{2n-4} \end{array}\   .  \label{eq:3.8}
\ee
and in addition we impose simple commutation relations between $M_n$ and
$V_{2n-2}$ and $V^{\prime}_{2n-1}$, respectively
\bse
\bea
M_{n+1,1} \cdot S^+_{12} V_{2n,2} &=& V_{2n,2} \cdot M_{n+1,1} \  ,
\label{3.9a} \\
V^{\prime}_{2n-1,2} S^-_{12}\cdot M_{n,1} &=& M_{n,1} \cdot V^{\prime}_{2n-1,
2}\   , \label{3.9b}
\eea
\ese
With the use of eqs. (3.8), (3.9) and (3.6) it is straightforward to derive
eqs. (3.1) and (3.3). Eq. (\ref{eq:3.4c}) can also be shown replacing $L_n$ by
$M^{-1}_{n+1} \, L^{\prime}_n \, M_n$ and taking account of the invariance of
commutation relations under the mapping. The relations (3.7)-(3.9) are
satisfied by the quantum mappings which will be considered in section 4. In
section 5 we construct commuting families of quantum invariants on the basis
of the full Yang-Baxter structure given above.

\section{The Quantum Lattice KdV and MKdV System}
\setcounter{equation}{0}

Here we consider two examples of integrable quantum mappings
coming from the lattice analogues of the KdV and MKdV equations.

\paragraph{a)}

The first example of a concrete integrable family of quantum mappings that
exhibit the structure outlined above, is the mapping of the KdV type (i.e.\@
mappings arising from the periodic initial value problem of lattice versions
of the KdV equation \cite{7} ). These are rational mappings ${\bf R}^{2P}
\rightarrow {\bf R}^{2P}: \left(\{v_j\}\right) \mapsto \left(\{v'_j\}\right)$
of the form
\be
v'_{2j-1}\ =\ v_{2j}   \    \ ,\     \  v'_{2j}\ =\ v_{2j+1}\ +\ \frac{\en
\dd}{v_{2j}}\ -\ \frac{\en \dd}{v_{2j+2}}  \qquad (j=1,\cdots ,
P),  \label{eq:xy}
\ee
imposing the periodicity condition $v_{i+2P}=v_i$. The mapping
(\ref{eq:xy}) has the Casimirs
\be
\sum_{j=1}^{P} v_{2j}\,=\,\sum_{j=1}^{P} v_{2j-1}\,=\,c \  , \label{eq:4.2}
\ee
where $c$ is chosen to be invariant under the mapping, in which case
we obtain a $(2P-2)$-dimensional generalization of the McMillan
mapping \cite{10}.

The mapping (\ref{eq:xy}) is obtained by reduction from the partial difference
equation
\be
u(n,m  + 1) - u(n+1,m) = q-p + \frac{p^2 - q^2}{p+q+u(n,m) - u(n+1,m+1)}
\label{eq:4.3}
\ee
for fields $u$ defined at the sites $(n,m)$ of a two-dimensional lattice, $n,m
\in {\bf Z}$. Eq. (\ref{eq:4.3}) is completely integrable in the sense that
solutions can be obtained via the direct linearization method, i.e.\@ by
solving a linear integral equation with arbitrary measure and contour
 \cite{6}.

To investigate eq. (\ref{eq:4.3}) one can choose initial data on a staircase
consisting of alternating horizontal and vertical steps, i.e.
\be
a_{2j} = u(j,j)\, , \, a_{2j+1} = u(j+1,j) \  , \label{eq:4.4}
\ee
and the solution above and below the staircase can be calculated from
(\ref{eq:4.3}) completing elementary squares. In the case of periodic initial
data
\be
a_j = a_{j+2P} \, , \, P \in {\bf N} \  , \label{eq:4.5}
\ee
the complete solution of (\ref{eq:4.3}) satisfies the periodicity property
\be
u(n,m) = u(n+P,m+P) \   . \label{eq:4.6}
\ee
The periodic solutions of (\ref{eq:4.3}) can be obtained via a
$2P$-dimensional mapping which is defined in terms of the vertical shift:
\be
u(n,m) \rightarrow u'(n,m) = u(n,m+1) \   . \label{eq:4.7}
\ee
In terms of the reduced variables
\be
v_j = p+q+a_j -a_{j+2} \label{eq:4.8}
\ee
the $2P$-dimensional mapping is given by eq. (\ref{eq:xy}).

To obtain the Yang-Baxter structure it is worthwhile to note that eq.
(\ref{eq:xy}) arises as the compatibility condition of a ZS system (1.1) with
\be
L_j\ =\ V_{2j} \cdot V_{2j-1}\   \ ,\   \
M_j\ =\ \left( \begin{array}{cc} u_j&1\\
\lambda_{2j}&0 \end{array} \right) \  ,  \label{eq:LM}
\ee
$$ V_i\ =\ \left( \begin{array}{cc} v_i&1\\
\lambda_i&0 \end{array} \right)
$$
in which $\lambda_{2j}=k^2-q^2$, $\lambda_{2j+1} = k^2 - p^2$ and $\en \dd
=p^2-q^2$. In fact, from the ZS condition (1.1) one obtains
\be
u_j \, = \, v_{2j-1} - \frac{\en \dd}{v_{2j}} \label{eq:4.10}
\ee
as well as the mapping (\ref{eq:xy}). The corresponding classical invariants,
obtained by expanding the trace of the monodromy matrix (1.6) in powers
of $k^2$, are in involution, cf.\@ eq. (1.9),  with respect to the
Poisson structure \cite{7}
\be
\{ v_j\, ,\,v_{j'} \}\ =\ \dd_{j+1,j'}\,-\,\dd_{j,j'+1} \  , \label{eq:xypb}
\ee
which was obtained using a Legendre transformation on an appropriately chosen
Lagrangian \cite{7}. This ensures that the mapping (\ref{eq:xy}) is
symplectic, i.e.\@ the same Poisson brackets hold also for the primes
variables $v'_j$. This property can also be checked easily by direct
computation. On the basis of this a canonical transformation to action-angle
variables can be found \cite{9}, thereby showing complete integrability in the
sense of Liouville \cite{7,8}. In the quantum case the variables $v_j$ become
hermitean operators on which we impose the following Heisenberg type of
commutation relations, as a natural quantization of the Poisson relations
(\ref{eq:xypb}), cf. \cite{14,21},
\be
\left[v_j\,,\,v_{j'} \right]\ =\ h\,\left(\dd_{j,j'+1}\,-
\,\dd_{j+1,j'}\right)\    ,  \label{eq:xycr}
\ee
(where $h=i\hbar$).
It is easy to check that the quantum mapping (\ref{eq:xy}) is a canonical
transformation with respect to these commutation relations.

The special solution of the quantum relations (2.2), (2.3), which
constitutes the $R,S$-matrix structure for the quantum mapping (\ref{eq:xy}),
together with the commutation relation (\ref{eq:xycr}), is given by
\bea
R^+_{12} & = & R^-_{12} - S^+_{12} + S^-_{12} \    \nn \\
R^-_{12} & = & {\bf 1}\otimes {\bf 1}\ +\
h\,\frac{P_{12}}{\mu_1\,-\,\mu_2}\  \label{eq:Rsol} \\
S^+_{12} & = & {\bf 1}\otimes {\bf 1}\ - \frac{h}{\mu_2} F \otimes E \  \  , \
 \  S^-_{12} = S^+_{21} \  , \nn
\eea
in which $\mu_{\alpha} = k^2_{\alpha} -q^2, \alpha = 1,2$ and the permutation
operator $P_{12}$ and the matrices $E$ and $F$ are given by
$$
P_{12} \ =\ \left( \begin{array}{cccc}
1&0&0&0\\0&0&1&0\\0&1&0&0\\0&0&0&1 \end{array} \right)
$$
\be
E\ =\ \left( \begin{array}{cc}
0&1\\0&0 \end{array} \right)\    \ ,\    \ F\ =\
\left( \begin{array}{cc}
0&0\\1&0 \end{array} \right) \  .     \label{eq:4.14}
\ee
We mention the useful identity
\be
R^+_{12}\ =\
\Lambda_1\Lambda_2\,R^-_{12}\Lambda_1^{-1}\Lambda_2^{-1}\ , \label{eq:4.15}
\ee
where $\Lambda_{\ar}=\mu_{\ar}F+E$, ($\ar=1,2$), from which it is evident that
it is not strictly necessary to introduce two different $R$-matrices
$R^{\pm}$.

The complete Yang-Baxter structure can now be derived from the mapping
(\ref{eq:xy}), the relation (\ref{eq:4.10}) for $u_j$ and the commutation
relation (\ref{eq:xycr}). In fact, from (\ref{eq:xycr}) one immediately
obtains eq. (\ref{eq:VV}) with
\be
S^+_{12} (n) = {\bf 1} \otimes {\bf 1} - \frac{h}{\lambda_{n,2}} F \otimes E
\label{4.16}
\ee
with $\lambda_{2j,2} = k^2_2 - q^2$, $\lambda_{2j-1,2} = k^2_2 - p^2$, and also
eqs. (\ref{eq:3.7b})  and (\ref{eq:3.7c}). These relations are at the basis of
the $L$ part, i.e.\@ eqs. (2.1), of the Yang-Baxter structure. To derive
the commutation relations (\ref{eq:3.8}), (3.9) one first checks by
explicit calculation that the only nonvanishing commutation relations between
the matrix $M_n$ and the matrices $V_m$, $V^{\prime}_m$ are indeed given by eq.
(\ref{eq:3.8}). Furthermore one has the commutation relations
\be
\left[ M_{n+1} - V_{2n+1} \stackrel{\otimes}{,} V_{2n} \right] = 0 \ \  , \ \
 \left[ M_{n+1} - V^{\prime}_{2n} \stackrel{\otimes}{,} V^{\prime}_{2n+1}
\right] = 0 \label{eq:4.17}
\ee
which with eq. (\ref{eq:VV}) and its counterpart in terms of the primed
operators immediately yield eqs. (3.9). Finally the nontrivial
commutation relations (3.2) follow from
\be
\left[ M_n \stackrel{\otimes}{,} M_n \right] = 0 \ \  , \  \  \left[
M^{\prime}_n - V^{\prime}_{2n-1} \stackrel{\otimes}{,} M_n \right] = 0
\ee
together with (3.9). The trivial commutation relations can be checked in
a similar way.

Thus, the mapping (\ref{eq:xy}) and its ZS system (\ref{eq:LM}) with the
commutation relation (\ref{eq:xycr}) satisfy the complete Yang-Baxter
structure treated in sections 2 and 3.\\

\noindent {\bf Remark:} The KdV mappings considered here are the discrete-time
analogue of the quantum Volterra system treated in ref. \cite{30}. Such
systems are of interest, in connection with discretizations of the Virasoro
algebra \cite{31}-\cite{34}.

\paragraph{b)}

We now consider the example of the MKdV mappings, which is associated
with the following $R,S$-matrice. Introducing
\be
R_{12}(x)\ =\ \left( \begin{array}{cccc}
qx-1&0&0&0\\ 0&x-1&q-1&0\\ 0&x(q-1)&q(x-1)&0\\0&0&0&qx-1
\end{array} \right)\        \ ,\       \
S_{12}\ =\ \left( \begin{array}{cccc}
1& & & \\  &1& & \\  & &1& \\ & & &q
\end{array} \right)\   \ ,
\ee
it is straightforward to check that the matrices of (4.19) obey for
spectral parameter $x = \lambda_1 / \lambda_2$ the following relations
\be
R_{12}\,\Lambda_1\,S_{21}\,\Lambda_2\ =\ \Lambda_2\,S_{12}\Lambda_1
\,R_{12}\        \  ,\       \
R_{12}\,\Lambda_1^{-1}\,S_{12}\,\Lambda_2^{-1}\ =\ \Lambda_2^{-1}\,
S_{21}\Lambda_1^{-1}\,R_{12}\   ,
\ee
in which $\Lambda_1 = \Lambda (\lambda_1 ) \otimes {\bf 1} \ \  . \ \
\Lambda_2 = {\bf 1} \otimes \Lambda (\lambda_2 )$ and
\be
\Lambda(\lambda )\ = \ \left( \begin{array}{cc} a&b\\ \lambda &d
\end{array} \right)\  ,
\ee
Eq. (4.20) then yields a solution of the Yang-Baxter relations (2.2),
(2.3) with
\bea
R^-_{12} & \equiv & R^-_{12} (\lambda_1 , \lambda_2 ) = R_{12} (\lambda_1
/ \lambda_2 ) \  \  ,\nn \\
R^+_{12} & = & \Lambda_1 \Lambda_2 R_{12}(\lambda_1 / \lambda_2 ) (\Lambda_1
\Lambda_2 )^{-1} =
R^-_{12} - S^+_{12} + S^-_{12} \  \   ,\nn       \\
S^+_{12} & = & \Lambda_2 S_{12} \Lambda^{-1}_2 \  \  , \nn   \\
S^-_{12} & = & S^+_{12} = \Lambda_1 S_{12} \Lambda^{-1}_1
\eea
As an example of a quantum mapping associated with this solution of the
Yang-Baxter equation we consider
\bea
\vf^{\prime}_{2n-1} &=& \vf_{2n}\    , \nn \\
e^{\vf^{\prime}_{2n}} &=&
\frac{ (p_{2n}-r)\,+\,(p_{2n+1}+r)e^{\vf_{2n+2}}}{ (p_{2n+1}-r)\,+\,
(p_{2n}+r)e^{\vf_{2n+2}}} \,e^{\vf_{2n+1}}\,
\frac{ (p_{2n-1}-r)\,+\,(p_{2n}+r)e^{\vf_{2n}}}{
(p_{2n}-r)\,+\,(p_{2n-1}+r)e^{\vf_{2n}}} \  ,   \nn \\
 & & n=1,\cdots 2P
\eea
On the classical level this mapping is obtained by reduction from the lattice
version of the MKdV equation \cite{6}:
\be
(p-r) \frac{w(n,m+1)}{w(n+1,m+1)} - (q-r) \frac{w(n+1,m)}{w(n+1,m+1)} =
(p+r) \frac{w(n+1,m)}{w(n,m)} - (q+r) \frac{w(n,m+1)}{w(n,m)} \label{4.24}
\ee
Choosing periodic initial values on a staircase of alternating horizontal and
vertical steps, i.e.
\be
b_{2j} = w(j,j) \, , \, b_{2j+1} = w(j+1,j) \, , \, b_{j+2P} = b_j \  \  ,
\label{4.25}
\ee
and defining the mapping in terms of the vertical shift, cf. (4.7) with $u
\rightarrow w$, we obtain eq. (4.23) in terms of the reduced variables
$\vf_j$ defined by
\be
e^{\vf_j} = \frac{b_j}{b_{j-2}}
\ee
The MKdV mapping (4.23) arises as the compatibility condition of a ZS system
(1.1) with $L_n = V_{2n}\cdot V_{2n-1}$, cf. (3.6), and
\[
V_n\ =\ \Lambda_n \cdot \overline{V}_n\   ,  \ \overline{V}_n\  = \ \left(
\begin{array}{cc} 1 &0\\ 0&e^{\vf_n} \end{array} \right)\   ,
\]
\be
\Lambda_n\ = \ \left( \begin{array}{cc} p_n-r& 1 \\ \lambda & p_n+r
\end{array} \right)\ .
\ee
in which $\lambda = k^2 - r^2$, $p_{2n-1} = p$, $p_{2n} = q$ and
\be
M_n\ =\ \Lambda_{2n} \cdot \left( \begin{array}{cc}
1&0\\ 0&e^{\gm_n} \end{array} \right)\   ,
\ee
In fact, working out (1.1) with (3.6) and (4.27), (4.28) one finds the mapping
(4.23) and
\be
e^{\gm_n}\ =\ \frac{ (p_{2n}-r)\,+\,(p_{2n-1}+r)e^{\vf_{2n}}}{(p_{2n-1}-r)\,
+\,(p_{2n}+r)e^{\vf_{2n}}}\,e^{\vf_{2n-1}}
\ee
At this stage it is worthwhile to note that the $v_j$ in eq. (4.27) and in eq.
(4.9) are related via a gauge transformation of the type
\be
\left( \begin{array}{cc} p_j -r & e^{\vf_j} \\ k^2-r^2 & (p_j +r) e^{\vf_j}
\end{array} \right) = \left( \begin{array}{cc} (p_j -r) b_{j-1} & b_j \\
(k^2-r^2) b_{j-1}  & 0 \end{array} \right) \; \left( \begin{array}{cc} v_j  &
1 \\ k^2-p^2_j & 0 \end{array} \right) \; \left( \begin{array}{cc} (p_{j-1}
-r) b_{j-2} & b_{j-1} \\ (k^2-r^2) b_{j-2}  & 0 \end{array} \right) ^{-1}
\ee
and the Miura transformation relating the KdV and MKdV mappings is given by
\be
v_j = (p_{j-1} -r) \frac{b_{j-2}}{b_{j-1}} + (p_j +r) \frac{b_j}{b_{j-1}}
\ee
In the classical case the mapping is completely integrable with $P$ integrals
in involution with respect to the (invariant) Poisson bracket
\be
\{ \vf_j , \vf_{j'} \} = \delta_{j',j+1} - \delta_{j',j-1}
\ee
cf. eq. (1.9) and the expansion of ${\rm tr} T(\lambda )$  in powers of $k^2$.

In the quantum case we have the commutation relation
\be
[\vf_j , \vf_{j'} ] = h \left( \delta_{j,j'+1} - \delta_{j,j'-1} \right)
\ee
implying in particular that
\be
e^{\vf_n} e^{\vf_{n+1}} = qe^{\vf_{n+1}} e^{\vf_n} \; , \; q = e^{-h}
\ee
Starting from (4.33) we find the commutation relations (3.7), in which the
$R^{\pm}_{12}, S^{\pm}_{12}$ are given by (4.22) with
\be
q = e^{-h} \; , \; x = \frac{k^2_1 - r^2}{k^2_2 - r^2}
\ee
and $\Lambda_n$ given by (4.27) with $\lambda = k^2 - r^2$.

 From the commutation relation (4.33), together with the explicit expression
(4.29) for $e^{\gamma_n}$ it is straightforward to derive the remaining
relations of the Yang-Baxter structure, i.e. eqs. (3.8), (3.9) and (3.2),
completely analogously to the case of the KdV-type of mappings. The trivial
commutation relations can also be checked directly. Thus with the MKdV-type of
mappings we have another example of the complete Yang-Baxter structure
presented in sections 2 and 3, but here the $R^{\pm}_{12}$ and $S^{\pm}_{12}$
correspond to different (trigonometric) solutions of the Yang-Baxter
equations.

\section{Quantum Invariants}
\setcounter{equation}{0}

In the classical case the trace of the monodromy matrix yields a sufficient
number of invariants which are in involution. In the quantum case the trace is
no longer invariant and we have to consider more general families of commuting
operators.

Following the treatment of ref. \cite{25}, a commuting parameter-family of
operators is obtained by taking (for details, cf. appendix D)
\be
\tau( \lambda )\ =\ tr\left( T(\lambda )K(\lambda )\right)\  ,     \label{5.1}
\ee
for any family of numerical matrices $K(\lambda )$ obeying the
relations
\be
K_1\,^{t_1\!}\left( (\,^{t_1\!}S^-_{12})^{-1}\right)
\,K_2\,R_{12}^+\ =\ R_{12}^-\,K_2\,^{t_2\!}\left(
(\,^{t_2\!}S^+_{12})^{-1}\right) \,K_1\   . \label{eq:RKSK}
\ee
(We assume throughout that $S_{12}^{\pm}$ and $R_{12}^{\pm}$ are
invertible). The left superscripts $\ ^{t_1}$ and $\ ^{t_2}$ denote
the matrix transpositions with respect to the corresponding factors
1 and 2 in the matricial tensor product.
Expanding (5.1) in powers of the spectral parameter $\lambda$, we obtain a set
of commuting observables of the quantum system in terms of which we can find a
common basis of eigenvectors in the associated Hilbert space. We note that a
matrix $K(\lambda )$ is commonly introduced in connection with quantum boundary
conditions other than periodic ones \cite{25}, but in relation to the quantum
mappings of the present paper it is essential in the periodic case as well.\\

Furthermore the Yang-Baxter equations of section 3 lead to the following
commutation relations between $M\equiv M_{n=1}$ and the monodromy matrix $T$,
\be
T_1\cdot M_1^{-1}\cdot S_{12}^+ M_2\ =\ M_2\cdot S_{12}^- T_1\cdot
M_1^{-1}   \  .  \label{5.3}
\ee
Here we use the notation $M_1 = M \otimes 1$, $M_2 = 1 \otimes M$ as usual for
the factors 1 and 2 in the matricial tensor product. The derivation of eq.
(5.3) is based on the commutation relation
\be
M_{n,1} \cdot \left( L_{n+1} \cdot L_n \cdot M^{-1}_n \right) _2 = \left(
L_{n+1} \cdot L_n \cdot M^{-1}_n \right) _2 \cdot S^-_{12} M_{n,1} \label{5.4}
\ee
which is easily checked noting that
\be
L_{n+1} \cdot L_n \cdot M^{-1}_n = M^{-1}_{n+2} \cdot L^{\prime}_{n+1} \cdot
L^{\prime}_n \label{5.5}
\ee
and using the commutation relation (3.1b) and the trivial relations (3.3b),
(3.3c). Then with the use of (3.1a) it is found that
\[
M_1 \cdot S^+_{12}  \left( L_P \cdot L_{P-1} \cdot \, \cdots \, \cdot L_1
\cdot M^{-1} \right) _2 = L_{P,2} \cdot M_1 \cdot \left( L_{P-1} \cdot \,
\cdots \, \cdot L_1 \cdot M^{-1} \right) _2
\]
\be
= \left( L_P \cdot \, \cdots \, \cdot L_3 \right) _2 \cdot \left( L_2 \cdot
L_1 \cdot M^{-1} \right) _2 \cdot S^-_{12} \cdot M_1
\ee
which is just eq. (5.3).

The commutation relation (2.8) for the monodromy matrices is invariant. This
can be shown noting that eqs. (2.1) are invariant under the mapping and
by repeating the derivation of eq. (2.8), but now with the updated variables
${L_j}^{\prime}$. It follows also directly from eq. (5.7) and the commutation
relation (5.3). In fact,
\bea
R^+_{12} T^{\prime}_1 \cdot S^+_{12} T^{\prime}_2 &=&
R^+_{12} M_1\cdot T_1\cdot M_1^{-1}\cdot S^+_{12} M_2\cdot
T_2\cdot M_2^{-1}  \nn \\
&=& R^+_{12} M_1\cdot M_2\cdot S^-_{12} T_1\cdot M_1^{-1}\cdot T_2\cdot
M_2^{-1}  \nn \\
&=& M_2\cdot M_1\cdot S^+_{12}R^+_{12} T_1\cdot M_1^{-1}\cdot T_2\cdot
M_2^{-1}  \nn \\
&=& M_2\cdot M_1\cdot S^+_{12}R^+_{12} T_1 \cdot S^+_{12} T_2 \cdot
M_2^{-1}\cdot M_1^{-1} S_{12}^{-^{-1}}  \nn \\
&=& M_2\cdot M_1\cdot S^-_{21} T_2\cdot S^-_{12} T_1\cdot R^-_{12}
M_2^{-1}\cdot M_1^{-1} S_{12}^{-^{-1}}  \nn \\
&=& T^{\prime}_2\cdot S_{21}^+ M_1\cdot M_2\cdot S_{12}^- T_1
M_1^{-1}\cdot M_2^{-1} S_{12}^{+^{-1}} R^-_{12}  \nn \\
&=& T_2^{\prime}\cdot S_{12}^- T_1^{\prime} R^-_{12}  \  ,
\eea

Our aim is now to describe the integrability of the quantum mappings of
section 4 which obey the commutation relations (2.8) and (5.3). For this we
need to show that one can find a sufficient family of commuting {\em
invariants} of the mapping. Let us thus use eq. (5.3) to calculate commuting
families of quantum invariants in the case of the KdV and MKdV mappings of
section 4.

In fact, introducing a tensor
\be
K_{12}\ =\ P_{12}K_1K_2\   ,
\ee
where $P_{12}$ is the permutation operator satisfying e.g.
\be
P_{12} A_1 = A_2 P_{12} \, , \, P_{12} A_2 = A_1 P_{12} \, , \, P_{12} =
P_{21} \, , \, tr_2 P_{12} = {\bf 1}_1
\ee
for matrices $A$ not depending on the spectral parameter, and choosing
$\lambda_1 = \lambda_2$, we can take the trace over left- and right hand side
of (\ref{5.3}) contracting with $K_{12}$. This leads to
\be
tr_{12}\left(K_{12}(TM^{-1})_1 \cdot S_{12}^+ M_2\right) \   =\ tr_{2}\left(
K_2 T_2
M_2^{-1}\,tr_1(P_{12}K_2S_{12}^+)\,M_2\right)\ =\ tr(KT)
\ee
provided that
\be
tr_1(P_{12}K_2S_{12}^+)={\bf 1}_2 \  .
\ee
In eq. (5.10) $tr_{12} = tr_1 tr_2$ denotes the trace over the factors 1 and 2
in the direct product space of matrices, whereas $tr_1$ and $tr_2$ are
restricted to only one of these factors. Under the same condition
(5.11) we have that
\be
tr_{12}\left(K_{12}M_2 \cdot S_{12}^- (TM^{-1})_1\right) \  =\ tr_{1}\left( K_1
M_1\,tr_2(P_{12}K_1S_{12}^-)\,(TM^{-1})_1\right)\ =\
tr(KT^{\prime})\  .
\ee
A common solution to eqs. (5.12) and (5.13)is easily found, namely
by taking
\be \label{eq:PS}
K_2\,=\,tr_1\left\{ P_{12} \,^{t_1\!}\left( ( \,^{t_1\!}S^+_{12})^{-1}
\right) \right\} \    .
\ee
It can be shown that (5.14) will solve eq. (\ref{eq:RKSK}).

Applying (5.14) to the examples of the KdV and MKdV mappings, we find
in the KdV case, using (4.13),
\be
K(\lambda ) = {\bf 1} + \frac{h}{\lambda} S_- \   \ ,\    \
S_- = \left( \begin{array}{cc}
0&0 \\ 0&1 \end{array} \right) \   ,
\ee
and in the case of the MKdV mapping, with the use of the relation
\be
S^-_{12} = {\bf 1} \otimes {\bf 1} + (q-1) \left( \Lambda_1 \cdot S_- \cdot
\Lambda_1^{-1} \otimes S_- \right) \  ,
\ee
cf. (4.19) and (4.20), we find the following solution from (4.22)
\bea
K(\lambda ) & = & 1 + (q^{-1} -1) S_- \cdot \Lambda \cdot S_- \cdot
\Lambda^{-1} \nn \\
\Lambda & = & \left( \begin{array}{cc} q-r & 1 \\ \lambda & q+r \end{array}
\right) \; , \; \lambda = k^2 - r^2
\eea
and again the $K(\lambda )$ in combination with the $R^{\pm}_{12} ,
S^{\pm}_{12}$ of eqs. (4.19), (4.20) satisfies eq. (5.2).

Hence, in the case of the KdV and MKdV mappings we have obtained a commuting
family of quantum invariants that can be evaluated expanding $\tau (\lambda )
= tr K(\lambda ) T(\lambda )$ in powers of $k^2$.

For instance, in the KdV mapping the explicit expression of the invariants can
be inferred from
\be
\tau(\lambda ) = :\,\left( \prod_{j=1}^{2P}\ v_j \right)
\left[ \,1\,+ \,
 \sum_{\stackrel{1\leq J_1<\cdots <J_N\leq 2P}{J_{\nu
+1}-J_{\nu}\geq 2,J_1-J_N+2P\geq 2}}
\prod_{\nu =1}^N\ \frac{\hat{\lambda}_{J_{\nu}}}{v_{J_{\nu}+1}v_{J_{\nu}}}
\right]\,: \  ,  \label{eq:TrT}
\ee
leading to find a full family of commuting invariants. In
(\ref{eq:TrT})\ :\,:\ denotes the normal ordering of the
operators $v_j$ in accordance with their enumeration, and
$\hat{\lambda}_J=\lambda_J$ for $J\neq 2P$, $\hat{\lambda}_{2P}=
\lambda_{2P}+h$. Thus the quantum effect is only visible in the boundary terms
associated with the factor $1/(v_1 v_{2P} )$.

As a very simple example we give the quantum invariant of the original
McMillan mapping \cite{10}, i.e. (\ref{eq:xy}) for $P=2$, namely
\be
x'\,=\,y\   \ ,\   \ y'\,=\,-x-\,\frac{2\en\dd y}{\en^2-y^2}\  ,
\label{eq:xymc}
\ee
for $x=v_1-\en$, $y=v_2-\en$, (choosing $c=2\en$) and
where $[y,x]=h$, having the invariant
\be
{\cal I}\ =\ (\en^2-y^2)(\en^2-x^2)\,-\,(\en\dd +h)( 2yx-\en\dd)\   .
\ee
The invariant ${\cal I}$ can be viewed as a Hamiltonian generating a
continuous-time interpolating flow by $\dot{x}=\frac{1}{h} [ {\cal I},x]$,
$\dot{y}=\frac{1}{h}[{\cal I},y]$, whose solutions can be considered to be
parametrized in terms of what we could call a quantum version of the Jacobi
elliptic functions. More general two-dimensional quantum mappings have been
studied in ref. \cite{35}.\\

\noindent {\bf Remark:} \\
The construction of quantum mappings can be generalized to a
larger class of models, namely those associated with the lattice
Gel'fand-Dikii hierarchy, \cite{36}, as was explicitly
shown in \cite{21}. In this case the $R,S$-matrix structure
is provided by the following solutions of the equations
(\ref{eq:RRR},\ref{eq:RSS}) and (2.3)
\bea
R^+_{12} & = & R^-_{12} - S^+_{12} + S^-_{12} = \Lambda_1 \Lambda_2 R^-_{12}
(\Lambda_1 \Lambda_2 )^{-1} \ \  , \nn \\
R^-_{12} & = &  {\bf 1}\ +\ h\,\frac{P_{12}}{\lambda_1\,- \,\lambda_2}\
   \ ,\     S^+_{12} = S^-_{21} = {\bf 1} - \frac{h}{\lambda_2}
\sum\limits^{N-1}_{i=1} E_{Ni} \otimes E_{iN}
\eea
\be
P_{12}\ =\ \sum_{i,j=1}^N\,E_{i,j}\otimes E_{j,i}\      \ ,
\      \  \Lambda(\ld)\ =\ \lambda E_{N,1}\,+\,\sum_{i=1}^{N-1}\,E_{i,i+1}\   ,
\ee
and $E_{i,j}$ being the basic generators of $Gl_N$, i.e. $(E_{i,
j})_{kl}=\dd_{ik}\dd_{jl}$. The relations of the full Yang-Baxter structure of
sections 2 and 3 are satisfied for matrices $L_n = V_{2n} \cdot V_{2n-1}$ and
$M_n$ of the form
\be
V_n\ =\ \Lambda_n \left( {\bf 1}\ +\ \sum_{i>j=1}^N\,v_{i,j}(n)E_{i,j}\
\right) \   ,  \   M_n = \Lambda_{2n} \left( {\bf 1} + \sum\limits^{N}_{i >
j=1} m_{i,j} (n) E_{i,j} \right)
\ee
in which $\Lambda_n=\Lambda(\ld_n)$, $\ld_{2n}=\ld_{2n+2}=\mu$ and the $v_{i,
j} (n)$ satisfy  the following Heisenberg type of commutation
relations, ($h=i\hbar$),
\be
\left[ v_{i,j}(n)\,,\,v_{k,l}(m) \right]\ =\
h\left( \dd_{n,m+1}\dd_{k,j+1}\dd_{i,N}
\dd_{l,1}\,-\,\dd_{m,n+1}\dd_{i,l+1}\dd_{k,N}\dd_{j,1} \right)\   .
\ee
A commuting family of quantum invariants is obtained from
\be
\tau (\lambda ) = {\rm tr} K(\lambda ) T(\lambda ) \ , \ K(\lambda ) = {\bf 1}
+ (N-1) \frac{h}{\lambda} E_{N,N}
\ee
but for $N \geq 3$ the expansion in powers of $k^2$ does not yield enough
invariants to establish complete integrability in the quantum case. For $N
\geq 3$ one needs additional invariants corresponding to higher order
commuting families of operators. The construction of these higher order
invariants needs the application of the fusion procedure \cite{29,37,38} and
is left to a future publication \cite{39}.

\pagebreak

\subsection*{Appendix A}
\setcounter{equation}{0}
\def\theequation{A.\arabic{equation}}

In order to derive the compatibility relations for the Yang-Baxter matrices
$R$ and $S$, i.e. eqs. (2.2), from the commutation relations (2.1) for the Lax
matrices $L$, we encounter four different types of combinations of matrices
$L$. Embedding the $L$ matrices in a tensorial product of three copies of the
matrix algebra, i.e. $L_n,j$, $j=1,2,3$ acting on vector spaces $V\otimes
V\otimes V$, and denoting
$$ L_{n,1}=L_n\otimes {\bf 1}\otimes {\bf 1}\   \ ,\   \
L_{n,2}={\bf 1}\otimes L_n\otimes {\bf 1}\   \ ,\   \
L_{n,3}={\bf 1}\otimes {\bf 1}\otimes L_n\  ,  $$
we can distinguish the following types of combinations of matrices $L$
involving only co\"{\i}nciding and/or neighbouring sites:

\begin{description}
\item[i)]\underline{$L_1\equiv L_{n+2,1}$, $L_2\equiv L_{n+1,2}$, $L_3
\equiv L_{n,3}$ }\\
\\
In this case, no conditions on the $R$- or $S$ matrices will appear,
because
\be
L_1\cdot S^+_{12}L_2\cdot S^+_{23}L_3\ =\ L_3\cdot L_2\cdot L_1\  ,
\label{ap1}
\ee
independently of the order in which the relation (\ref{eq:RLL}) is
applied.

\item[ii)]\underline{$L_1\equiv L_{n+1,1}$, $L_2\equiv L_{n+1,2}$,
$L_3\equiv L_{n,3}$ }\\
\\
In this case, we have on the one hand
\be
R^+_{12}L_1\cdot L_2\cdot S^+_{13}S^+_{23}L_3\,=\,L_2\cdot L_1\cdot
R^-_{12}S^+_{13}S^+_{23}L_3\ ,  \label{eq:B1}
\ee
whereas on the other hand we find
\bea
R^+_{12}L_1\cdot S^+_{13}L_2\cdot S^+_{23}L_3 &=& R^+_{12}L_3\cdot
L_1\cdot L_2\  \nn \\
=\,L_3\cdot L_2\cdot L_1R^-_{12} &=& L_2\cdot S^+_{23}L_1\cdot S^+_{13}L_3
R^-_{12}\  .  \label{eq:B2}
\eea
Comparing  relations  (\ref{eq:B1}) and (\ref{eq:B2}), we have
\be
R^-_{12}S^+_{13}S^+_{23}\,=\,S^+_{23}S^+_{13}R^-_{12}\  , \label{apA4}
\ee
which after relabelling  of the vector spaces becomes eq. (2.2b).

\item[iii)]\underline{$L_1\equiv L_{n+1,1}$, $L_2\equiv L_{n,2}$,
$L_3\equiv L_{n,3}$ }\\

Take for this case the combination
\be
R^+_{23}L_1\cdot S^+_{12}L_2\cdot S^+_{13}L_3\,=\,
L_1\cdot R^+_{23}S^+_{12}S^+_{13}\cdot L_2 \cdot L_3\  ,  \label{eq:B3}
\ee
and compare this with
\bea
R^+_{23}L_1\cdot S^+_{12}L_2\cdot S^+_{13}L_3 &=&
R^+_{23}L_2\cdot L_3\cdot L_1  \nn \\
&=& L_3\cdot L_2\cdot L_1 R^-_{23}  \nn \\
&=& L_1\cdot S^+_{13}S^+_{12}L_3\cdot L_2R^-_{23} \label{eq:B4} \\
&=& L_1\cdot S^+_{13}S^+_{12}R^+_{23} L_2\cdot L_3\  ,  \nn
\eea
yielding eq. (2.2b) with the + sign.

\item[iv)]\underline{$L_1\equiv L_{n,1}$, $L_2\equiv L_{n,2}$,
$L_3\equiv L_{n,3}$ }\\
\\
In this case we have the standard braiding type of argument to find as a
sufficient condition the quantum $R$-matrix relations (2.2a) for $R^+$ and
$R^-$.
\end{description}

\subsection*{Appendix B}
\setcounter{equation}{0}
\def\theequation{B.\arabic{equation}}

In this appendix we establish the commutation relations between the monodromy
matrices $T$ and $T^+_n,T^-_n$ of eq. (2.6), using the fundamental commutation
relations of the matrices $L_n$. \\

\noindent
{\bf i)} Using eqs. (2.1a), (2.1b) we can establish
\bea
R_{12}^+\,L_{n+1,1} \cdot L_{n,1} \cdot L_{n+1,2} \cdot L_{n,2} &=&
R_{12}^+\,L_{n+1,1} \cdot \,L_{n+1,2} \cdot S^+_{21}\,L_{n,1} \cdot
L_{n,2} \nn \\
&=& \,L_{n+1,2} \cdot \,L_{n+1,1} \cdot R_{12}^-\,S^+_{21}
\,L_{n,1} \cdot \,L_{n,2} \  , \label{apB1}
\eea
which, by imposing  the relation (2.3),  reduces to
\bea
\                                       \ &=& \,L_{n+1,2} \cdot
L_{n+1,1} \cdot
S^+_{12} \,L_{n,2} \cdot L_{n,1}\, R_{12}^- \  , \nn \\
&=& L_{n+1,2} \cdot L_{n,2} \cdot L_{n+1,1} \cdot
L_{n,1}\,R_{12}^- \  . \label{eq:C1}
\eea
By repeated application of eqs. (\ref{apB1}) and (\ref{eq:C1}) together with
eq. (2.3) one shows that
\bea
&R^+_{12}&L_{P,1}\cdot \dots \cdot L_{n+1,1}\cdot L_{P,2}\cdot \dots
\cdot L_{n+1,2} \nn \\
&=& R^+_{12}\,L_{P,1}\cdot L_{P-1,1}\cdot L_{P,2}\cdot L_{P-2,1}
\cdot L_{P-1,2}\cdot \dots \cdot L_{n+1,1}\cdot L_{n+2,2}
\cdot L_{n+1,2} \nn \\
&=& R^+_{12}\,L_{P,1}\cdot L_{P,2}\cdot S^+_{21}L_{P-1,1}\cdot L_{P-1,2}
\cdot S^+_{21}\cdot L_{P-2,1}\cdot \dots \cdot L_{n+2,2}\cdot S^+_{21}
L_{n+1,1}\cdot L_{n+1,2} \nn \\
&=& L_{P,2}\cdot L_{P,1}S^+_{12}\cdot L_{P-1,2}\cdot
L_{P-1,1}S^+_{12} \cdot L_{P-2,2}\cdot \dots \cdot L_{n+2,1}S^+_{12}\cdot
L_{n+1,2}\cdot L_{n+1,1}\,R^-_{12} \nn \\
&=& L_{P,2}\cdot L_{P-1,2}\cdot \dots \cdot L_{n+1,2} \cdot L_{P,1}
\cdot L_{P-1,1}\cdot \dots \cdot L_{n+1,1}\,R^-_{12}\   ,  \label{eq:C2}
\eea
leading to eq. (2.7a) for $T^+_n$. A similar argument can be applied for
$T^-_n$. Furthermore, eq. (2.7b) is derived from eqs. (2.1b), (2.1c) by noting
that
\bea
&L_{P,1}&\cdot \dots \cdot L_{n+1,1}\cdot S^+_{12}L_{n,2}\cdot \dots
\cdot L_{1,2}\nn \\
&=& L_{P,1}\cdot \dots \cdot L_{n+2,1}\cdot L_{n,2}\cdot L_{n+1,1}
\cdot L_{n-1,2} \cdot \dots \cdot L_{1,2}\nn \\
&=& L_{n,2}\cdot \dots \cdot L_{2,2}\cdot L_{P,1}\cdot L_{1,2}\cdot
L_{P-1,1} \cdot \dots \cdot L_{n+1,1}\nn \\
&=& L_{n,2}\cdot \dots \cdot L_{2,2}\cdot L_{1,2}S^+_{21}L_{P,1}\cdot
\dots \cdot L_{n+1,1}\ , \label{eq:C3}
\eea
where in the last step we have used the commutation relation
\be
L_{P,1}\cdot L_{1,2}\ =\ L_{1,2}\cdot S^+_{21} L_{P,1} \label{apB5}
\ee
taking into account the periodic boundary conditions.

Finally, from (2.7) we immediately obtain
\bea
R^+_{12}\,T_1\cdot S^+_{12} T_2 &=& R^+_{12} T_{n,1}^+\cdot T_{n,2}^+
\cdot S^+_{21}T_{n,1}^-\cdot T_{n,2}^- \nn \\
&=&  T_{n,2}^+\cdot T_{n,1}^+\cdot  R^-_{12}S^+_{21}T_{n,1}^-\cdot
T_{n,2}^- \nn \\
&=& T_{n,2}^+\cdot T_{n,1}^+\cdot S^+_{12}T_{n,2}^-\cdot T_{n,1}^-
R^-_{12} \nn  \\
&=& T_2\cdot S^+_{21} T_1 R^-_{12} \  , \label{eq:C4}
\eea
which is eq. (2.8).

\subsection*{Appendix C}
\setcounter{equation}{0}
\def\theequation{C.\arabic{equation}}

{\bf i)} We prove that eqs. (3.1a),(3.1b), together with (3.2b) are sufficient
to ensure that the basic commutation relations between the matrices $L_n$,
eqs. (2.1a), are preserved under the mapping
$$ L_n\mapsto L_n^{\prime}\,=\,M_{n+1}\cdot L_n\cdot M_n^{-1}. $$
In fact,
\bea
L^{\prime}_{n+1,1} \cdot S^+_{12} L^{\prime}_{n,2} &=&
L^{\prime}_{n+1,1} \cdot S^+_{12} M_{n+1,2} \cdot L_{n,2} \cdot M^{-1}_{n,2}
\nn \\
&=& M_{n+1,2} \cdot L^{\prime}_{n+1,1} \cdot L_{n,2} \cdot M^{-1}_{n,2} \nn \\
&=& M_{n+1,2} \cdot M_{n+2,1} \cdot L_{n+1,1} \cdot M_{n+1,1}^{-1} \cdot
L_{n,2} \cdot M_{n,2}^{-1} \nn \\
&=& M_{n+1,2} \cdot M_{n+2,1} \cdot L_{n+1,1} S^+_{12} L_{n,2} \cdot
M_{n+1,1}^{-1}  \cdot M_{n,2}^{-1} \nn \\
&=& L^{\prime}_{n,2} \cdot M_{n,2} \cdot L^{\prime}_{n+1,1} \cdot
M_{n,2}^{-1} \nn \\
&=& L^{\prime}_{n,2} \cdot L^{\prime}_{n+1,1}\  ,
\eea
and similarly
\bea
R^+_{12} L^{\prime}_{n,1} \cdot L^{\prime}_{n,2} &=& R^+_{12} M_{n+1,1}
\cdot L_{n,1} \cdot M^{-1}_{n,1} \cdot L^{\prime}_{n,2} \nn \\
&=& R^+_{12} M_{n+1,1}\cdot L_{n,1} \cdot L^{\prime}_{n,2} \cdot
M^{-1}_{n,1} (S^-_{12})^{-1} \nn \\
&=& R^+_{12} M_{n+1,1} \cdot M_{n+1,2} \cdot S^-_{12} L_{n,1} \cdot
L_{n,2} \cdot M^{-1}_{n,2} \cdot M^{-1}_{n,1} (S^-_{12})^{-1} \nn \\
&=& M_{n+1,2} \cdot M_{n+1,1} \cdot S^+_{12} R^+_{12} L_{n,1} \cdot
L_{n,2} \cdot M^{-1}_{n,2} \cdot M^{-1}_{n,1} (S^-_{12})^{-1} \nn \\
&=& M_{n+1,2} \cdot M_{n+1,1} \cdot S^+_{12} L_{n,2} \cdot L_{n,1} \cdot
M^{-1}_{n,1} \cdot M^{-1}_{n,2} (S^+_{12})^{-1} R^-_{12} \nn \\
&=& M_{n+1,2} \cdot L_{n,2} \cdot M_{n+1,1} \cdot L_{n,1} \cdot
M^{-1}_{n,1} \cdot M^{-1}_{n,2} (S^+_{12})^{-1} R^-_{12} \nn \\
&=& L^{\prime}_{n,2} \cdot M_{n,2} \cdot L^{\prime}_{n,1} \cdot
M^{-1}_{n,2} (S^+_{12})^{-1} R^-_{12} \nn \\
&=& L^{\prime}_{n,2} \cdot L^{\prime}_{n,1} R^-_{12} \   .
\eea
Finally, we have that
\bea
M^{\prime}_{n+1,1} \cdot S^+_{12} L^{\prime}_{n,2} &=&
M^{\prime}_{n+1,1} \cdot S^+_{12} M_{n+1,2} \cdot L_{n,2} \cdot
M^{-1}_{n,2}  \nn \\
&=& M_{n+1,2} \cdot M^{\prime}_{n+1,1} \cdot L_{n,2} \cdot M^{-1}_{n,2}  \nn \\
&=& M_{n+1,2} \cdot L_{n,2} \cdot M^{\prime}_{n+1,1} \cdot M^{-1}_{n,2}  \nn \\
&=& L^{\prime}_{n,2} \cdot M^{\prime}_{n+1,1}\     .
\eea
It is straightforward to check along similar lines that all trivial
commutation relations remain so after applying the mapping.\\

\subsection*{Appendix D}
\setcounter{equation}{0}
\def\theequation{D.\arabic{equation}}

In order to show that eq. (5.2) provides a commuting family of operators, let
us give an argument similar to the one given by Sklyanin in \cite{25}.
Denoting by $T_i$ and $K_i$, ($i=1,2$) the monodromy matrix resp. matrix $K$
for two different values of the spectral parameter $\lambda_1$ resp.
$\lambda_2$ and acting in two different factors of a matricial tensor product,
and denoting by $\tau_1$,$\tau_2$ the invariants (5.1) evaluated at these
respective values of the spectral parameter,
we have on the one hand, assuming that $[K_i \stackrel{\otimes}{,} T_j ] = 0$,
\bea
\tau_1 \, \tau_2
&=& tr_1\left( T_1K_1\right) \,
tr_2\left( T_2K_2\right)\
=\ tr_{1,2}\left\{ T_1K_1\, \,^{t_2\!}T_2\,^{t_2\!}K_2\right\} \nn \\
&=& tr_{1,2}\left\{ \,^{t_2\!}(T_1\, S^+_{1,2}T_2)\ ^{t_1\!}(
\,^{t_1\!}K_1\, \,^{t_{1}\!}\left( (\,^{t_2\!}S^+_{1,2})^{-1}\right)
\,^{t_2\!}K_2)\,\right\} \nn \\
&=& tr_{1,2}\left\{ R_{1,2}^{+^{-1}}\,T_2\, S^-_{1,2}T_1\,R_{1,2}^-\
^{t_{12}\!}\left[ \,^{t_1\!}K_1\, \,^{t_{1}\!}\left(
(\,^{t_2\!}S^+_{1,2})^{-1}\right) \,^{t_2\!}K_2 \right]\,\right\}\
\label{eq:D1}\\
&=& tr_{1,2}\left\{ \,T_2\, S^-_{1,2}T_1\ ^{t_{12}\!}\left[
(\,^{t_{12}\!}R^+_{1,2})^{-1}
\,^{t_1\!}K_1\, \,^{t_{1}\!}\left( (\,^{t_2\!}S^+_{1,2})^{-1}\right)
\,^{t_2\!}K_2 \,^{t_{12}\!}R_{1,2}^- \right] \right\}\  ,
\nn \eea
whereas on the other hand we have
\bea
\tau_2 \, \tau_1
&=& tr_2\left( T_2K_2\right) \,
tr_1\left( T_1K_1\right) \nn \\
&=& tr_{1,2}\left\{ \,^{t_1\!}\left( T_2\,
S^-_{1,2}T_1\right)\ ^{t_2\!}\left[ \,
^{t_2\!}K_2\, \,^{t_2\!}\left( (\,^{t_1\!}S^-_{1,2})^{-1}\right)
\,^{t_1\!}K_1\right] \right\}
\label{eq:D2} \\
&=& tr_{1,2}\left\{ T_2\, S^-_{1,2}T_1\ ^{t_{12}\!}(
\,^{t_2\!}K_2\, \,^{t_{2}\!}\left( (\,^{t_1\!}S^-_{1,2})^{-1}\right)
\,^{t_1\!}K_1)\,\right\} \    , \nn
\eea
from which it is clear that (D.1) and (D.2) can be identified provided that we
have the following condition on the matrices $K$
\be
\left( \,^{t_{12}\!}R^+_{12}\right)^{-1} \,^{t_1\!}K_1\,^{t_1\!}\left(
(\,^{t_2\!}S^+_{12})^{-1}\right) \,^{t_2\!}K_2\ =\
\,^{t_2\!}K_2\,^{t_2\!}\left(
(\,^{t_1\!}S^-_{12})^{-1}\right) \,^{t_1\!}K_1\,
\left( \,^{t_{12}\!}R^-_{12}\right)^{-1}\    . \label{eq:D3}
\ee
Eq. (D.3) is a very general condition for operator valued matrices $K$ of
which the entries commute with the entries of $T$, which is sufficient to
ensure that the $T(\lambda )$ form a parameter family of commuting operators.
For numerical matrices $K(\lambda )$ eq. (D.3) leads to the condition (5.2)
given in the main text.


\begin{thebibliography}{99}
\bibitem{1}
V.G. Drinfel'd, {\em Quantum Groups}, Proc. ICM Berkeley 1986, ed.
A.M. Gleason, (AMS, Providence, 1987), p. 798.
\bibitem{2}
M. Jimbo, Lett. Math.
Phys. {\bf 10} (1985) 63, ibid. {\bf 11} (1986) 247, Commun. Math. Phys.
{\bf 102} (1986) 537.
\bibitem{3}
L.D. Faddeev, N.Yu. Reshetikhin and L.A. Takhtadzhyan, Algebra i Analiz.
{\bf 1} (1988) 178 [in Russian].
\bibitem{4}
L.D. Faddeev, in {\em D\'eveloppements R\'ecents en Th\'eorie des
Champs et M\'ecanique Statistique}, eds. J.-B. Zuber and R. Stora,
(North-Holland Publ. Co., 1984), p.561.
\bibitem{5}
M.J. Ablowitz and F.J. Ladik, Stud. Appl. Math. {\bf 55} (1976)
213, {\bf 57} (1977) 1;
R. Hirota, J. Phys. Soc. Japan {\bf 43} (1977) 1424, 2074, 2079;
{\bf 50} (1981) 3785;
E. Date, M. Jimbo and T. Miwa, J. Phys. Soc. Japan {\bf 51}
(1982) 4116, 4125, {\bf 52} (1983) 388, 761, 766.
\bibitem{6}
F.W. Nijhoff, G.R.W. Quispel and H.W. Capel, Phys. lett. {\bf 97A}
(1983) 125;
G.R.W. Quispel, F.W. Nijhoff, H.W. Capel and J. van der Linden, Physica
{\bf 125A} (1984) 344.
\bibitem{7}
V.G. Papageorgiou, F.W. Nijhoff and H.W. Capel, Phys. Lett. {\bf 147A} (1990)
106;
H.W. Capel, F.W. Nijhoff and V.G. Papageorgiou, Phys. Lett. {\bf 155A}
(1991) 377.
\bibitem{8}
F.W. Nijhoff, V.G. Papageorgiou and H.W. Capel, {\em Integrable
Time-Discrete Systems: Lattices and Mappings}, in Proc. of the Intl.
Workshop on Quantum Groups, The Euler Intl. Math. institute, Leningrad,
ed. P.P. Kulish, Springer Lecture Notes Math. {\bf 1510} (1992) 312.
\bibitem{9}
M. Bruschi, O. Ragnisco, P.M. Santini and G.-Z. Tu, Physica {\bf 49D}
(1991) 273.
\bibitem{10}
E.M. McMillan, in {\em Topics in Physics}, eds. W.E. Brittin and H. Odabasi,
(Colorado Associated Univ. Press, Boulder, 1971), p. 219.
\bibitem{11}
G.R.W Quispel, J.A.G. Roberts and C.J. Thompson, Phys. Lett. {\bf A126}
(1988) 419, Physica {\bf D34} (1989) 183.
\bibitem{12}
A.P. Veselov, Funct. Anal. Appl. {\bf 22} (1988) 83; Theor. Math. Phys.
{\bf 71} (1987) 446;
P.A. Deift and L.C. Li, Commun. Pure Appl. Math. {\bf 42} (1989) 963.
J. Moser and A.P. Veselov, Preprint ETH (Z\"urich), 1989.
\bibitem{13}
Yu.B. Suris, Phys. Lett. {\bf 145A} (1990) 113; Algebra i Analiz {\bf 2} (1990)
141 [in Russian].
\bibitem{14}
F.W. Nijhoff, H.W. Capel and V.G. Papageorgiou, Phys. Rev. {\bf 46A}
(1992) 2155.
\bibitem{15}
L.D. Faddeev and L.A. Takhtadzhyan, {\em Hamiltonian Methods in the
Theory of Solitons}, (Springer Verlag, Berlin, 1987).
\bibitem{16}
J.M. Maillet, Phys. Lett. {\bf 162B} (1985) 137, Nucl. Phys.
{\bf B269} (1986) 54.
\bibitem{17}
A.G. Reyman and M.A. Semenov-Tian-Shanskii, Phys. Lett. {\bf 130A}
(1988) 456.
\bibitem{18}
O. Babelon and C. Viallet, Phys. Lett. {\bf 237B} (1990) 411.
\bibitem{19}
J. Avan and M. Talon, Nucl. Phys. {\bf B352} (1991) 215.
\bibitem{20}
L.-C. Li and S. Parmentier, C.R. Acad. Sci. Paris, t. {\bf 307}
S\'erie I (1988) 279; Commun. Math. Phys. {\bf 125} (1989) 545.
\bibitem{21}
F.W. Nijhoff and H.W. Capel, Phys. Lett. {\bf 163A} (1992) 49.
\bibitem{22}
O. Babelon and L. Bonora, Phys. Lett. {\bf 253B} (1991) 365;
O. Babelon, Commun. Math. Phys. {\bf 139} (1991) 619. .
\bibitem{23}
A. Alekseev, L.D. Faddeev and M.A. Semenov-Tian-Shanskii, Peprint
CERN-TH-5981/91.
\bibitem{24}
A. Alekseev, L.D. Faddeev and M.A. Semenov-Tian-Shanskii,
in Proc. of the Intl.
Workshop on Quantum Groups, The Euler Intl. Math. institute, Leningrad,
ed. P.P. Kulish, Springer Lecture Notes Math. {\bf 1510} (1992) 148.
\bibitem{25}
E.K. Sklyanin, J. Phys. {\bf A21} (1988) 2375.
\bibitem{26}
N. Yu. Reshetikhin and M.A. Semenov-Tian-Shanskii, Lett. Math. Phys.
{\bf 19} (1990) 133.
\bibitem{27}
G.I. Olshanskii, {\em Twisted Yangians and infinite-dimensional classical
Lie algebras}, Preprint CWI (Amsterdam), 1990.
\bibitem{28}
A. Alekseev and L.D. Faddeev, Commun. Math. Phys. {\bf 141} (1991) 413.
\bibitem{29}
A.G. Izergin and V.E. Korepin, Sov. Phys. Dokl. {\bf 26} (1981) 653.
\bibitem{30}
A.Yu. Volkov, Preprint HU-TFT-92-6.
\bibitem{31}
J.L. Gervais, Phys. Lett. {\bf 160B} (1985) 277,279.
\bibitem{32}
L.D. Faddeev and L.A. Takhtadzhyan, Springer Lect. Notes Phys. {\bf 246}
(1986) 166.
\bibitem{33}
A.Yu. Volkov, Theor. Math. Phys. {\bf 74} (1988) 96.
\bibitem{34}
O. Babelon, Phys. Lett. {\bf 215B} (1988) 523, ibid. {\bf 238B} (1990)
234.
\bibitem{35}
G.R.W. Quispel and F.W. Nijhoff, Phys. Lett. {\bf 161A} (1991)
419.
\bibitem{36}
F.W. Nijhoff, H.W. Capel, G.R.W. Quispel and V.G. Papageorgiou,
Inverse Probl. {\bf 8} (1992) 597.
\bibitem{37}
P.P. Kulish and E.K. Sklyanin, Springer Lect. Notes Phys. {\bf 151}
(1982) 61.
\bibitem{38}
P.P. Kulish, N. Yu. Reshetikhin and E.K. Sklyanin, Lett. Math.
Phys. {\bf 5} (1981) 393.
\bibitem{39}
F.W. Nijhoff and H.W. Capel, {\em Integrability and Fusion Algebra for Quantum
Mappings}, in preparation.
\end{thebibliography}
\end{document}